\documentclass[pra,twocolumn,showpacs,floatfix,superscriptaddress,aps]{revtex4}
\usepackage{amsmath}
\usepackage{makeidx}
\usepackage{amssymb}
\usepackage{graphicx}
\usepackage[usenames]{color}

\newcommand{\beq}{\begin{equation}}
\newcommand{\eeq}{\end{equation}} 
\newcommand{\beqa}{\begin{eqnarray}}
\newcommand{\eeqa}{\end{eqnarray}}
\newcommand{\ba}{\begin{array}}
\newcommand{\ea}{\end{array}}

\begin{document}

\title{Dimensional reduction of a binary Bose-Einstein condensate 
in mixed dimensions}
\author{L. E. Young S.$^{1}$\footnote{lyoung@ift.unesp.br}, 
L. Salasnich$^{2,3}$\footnote{luca.salasnich@cnr.it}, 
and S. K. Adhikari\footnote{adhikari44@yahoo.com;
http://www.ift.unesp.br/users/adhikari/}}
\affiliation{Instituto de F\'isica Te\'orica, 
UNESP $-$ Universidade Estadual
Paulista,01.140-070 S\~ao Paulo, S\~ao Paulo, Brazil \\
$^2$INO-CNR, Research Unit at the Dipartimento di Fisica ``Galileo
Galilei'', Universit\`a di Padova, Via Marzolo 8, 
35131 Padova, Italy\\
$^{3}$CAMTP, University of Maribor, Krekova 2, 2000 Maribor, Slovenia}

\begin{abstract}

We present effective reduced equations for the study of a binary 
Bose-Einstein condensate (BEC), where the confining potentials of the 
two BEC components have distinct asymmetry so that the components belong 
to different space dimensions as in a recent experiment [G. Lamporesi 
{\it et al.}, Phys. Rev. Lett. {\bf 104,} 153202 (2010)]. Starting from 
a binary three-dimensional (3D) Gross-Pitaevskii equation (GPE) and 
using a Lagrangian variational approach we derive a binary effective 
nonlinear Schr\"odinger equation with components in different reduced 
dimensions, e. g., the first component in one dimension and the second 
in two dimensions as appropriate to represent a cigar-shaped BEC coupled 
to a disk-shaped BEC. We demonstrate that the effective reduced 
binary equation, which depend on the geometry of the system, is quite 
reliable when compared with the binary 3D GPE and can be efficiently 
used to perform numerical simulation and analytical calculation for the 
investigation of static and dynamic properties of 
a binary BEC in mixed dimensions.

\end{abstract}


\maketitle

\section{Introduction}

It is now well established that, at ultralow temperature, 
dilute Bose-Einstein condensates (BECs) 
can be accurately described by the three-dimensional (3D) Gross-Pitaevskii 
equation (GPE) \cite{stringari,leggett}. 
Some years ago it has been found that, starting from the 3D GPE 
and using a Gaussian variational approach \cite{PerezGarcia98,kav}, 
it is possible to derive effective one-dimensional
(1D) or two-dimensional 
(2D) wave equations 
which describes the dynamics of quasi-1D or quasi-2D BECs  
as appropriate for cigar or disk shapes
\cite{sala1}. For instance, in that derivation of the effective 
1D equation the trapping potential is 
harmonic and isotropic in the transverse direction 
and generic in the axial one. The effective 
1D equation, which is a time-dependent 
nonpolynomial Schr\"odinger equation (NPSE) \cite{sala1}, 
has been used to model a cigar-shaped condensate by many experimental 
and theoretical groups \cite{toscani,tedeschi,spagnoli,camilo,oth}. 
Similar reduction scheme also exists for a disk-shaped 
BEC leading to a 2D NPSE.
(The lowest-order approximation of the NPSE leads to a GP-type 
nonlinear Schr\"odinger equation (NLSE)
equation with cubic nonlinearity.
Different variants of this reduction scheme have later been suggested,
which offer certain advantages in some cases \cite{spagnoli,camilo}.)
Moreover, by using 
the 1D NPSE,  analytical and numerical solutions 
have been found 
for solitons and vortices \cite{sala3}. Recently,
a discrete version of the 1D NPSE (1D DNPSE) has been obtained 
\cite{sala4}. 
The 1D DNPSE has been used to model a BEC confined in a combination of 
a cigar-shaped trap and a deep optical lattice acting in the axial 
direction \cite{sala4}. More recently the 2D NPSE, obtained for a 
disk-shaped BEC subject to strong confinement along the axial 
direction, has been used to predict the density profiles and dynamical 
stability of repulsive and attractive BECs with zero and finite 
topological charge in various planar trapping configurations, including 
the axisymmetric harmonic confinement and 1D periodic potential 
\cite{sala5}. The reduction scheme has also 
successfully been applied \cite{camilo,camilo2} 
to the study of a density-functional formulation  \cite{df}
for a Fermi superfluid at unitarity as well as to the study of 
Anderson localization in a BEC \cite{ander}.

In a  very recent experiment Lamporesi {\it et al.} 
\cite{inguscio} investigated 
the mixed-dimensional scattering 
occurring when the collisional 
atoms of two species of a binary BEC 
live in different space dimensions. This achievement 
opens the way to the experimental and theoretical 
studies of a binary BEC with the components belonging 
 to mixed dimensions. 
In the present paper we consider the dimensional reduction 
of a binary 3D GPE appropriate for the study of 
a binary BEC in mixed dimensions. 
Starting from a binary 3D 
GPE we derive a binary nonpolynomial 
Schr\"odinger equation (BNPSE) in reduced 
dimensions by using a Lagrangian variational 
approach. (The BNPSE represent a set of  coupled nonlinear differential 
equations with 
nonpolynomial non-power-law nonlinearities.)
We also present a simplified version of 
the reduction scheme first, where the reduced equations in the form of a 
binary NLSE (BNLSE)
have simpler power-law
 nonlinearities.
The BNPSEs depend on the geometry of the system 
and also on the inter-species interaction. 
In fact, the main difference with respect to the case 
of a single BEC is  the coupling between the two BECs, 
which gives rise, quite generally, to a integral term 
in the reduced wave equations.  
(In the past, a binary  1D NPSE has been derived and 
used \cite{sala6} to study a binary cigar-shaped BEC, rather than a
binary BEC in mixed dimensions as considered here.) 

We use the BNLSE and BNPSE to study numerically the static and dynamic 
properties of a 1D-2D binary BEC where the first quasi-1D component lies 
along the $z$ axis and the quasi-2D component in the $(x,y)$ plane. For 
small nonlinearities both the BNLSE and BNPSE yielded results for 
density in good agreement with the full 3D result. At larger values of 
nonlinearities the BNPSE provided better approximation to the full 3D 
result. We also studied small oscillations of the  1D-2D binary BEC initiated 
by a sudden change of the radial angular frequency of the 
quasi-2D component. The  BNPSE 
provided a better description of this oscillation over the  BNLSE, when 
compared with the full 3D model, specially for large nonlinearities and 
large times. 

The  reduced effective BNLSEs 
with simple power-law 
nonlinearity, but with different dimensionalities for the components, 
are presented in Sec. \ref{II}. These equations are derived 
by considering the ansatz for the wave function that sets the transverse 
spatial dependence as the Gaussian ground states in transverse harmonic 
traps. The transverse spatial dependence is finally integrated out and 
the reduced equations derived in the usual fashion using a Gaussian 
variational approach. In Sec. \ref{III} 
the BNPSEs with different dimensionalities for the components are 
derived by taking the ansatz for the wave function that sets the 
transverse spatial dependence as generic Gaussian states which 
are not the ground states in transverse harmonic traps. The reduced effective 
BNPSEs are then derived by a Lagrangian variational approximation which
determines the widths of the Gaussian functions in a self-consistent fashion. 
In Sec. \ref{IV} we present the numerical results, where we compare the 
densities and chemical potentials of the 3D binary GPE with those of the 
BNLSE and BNPSE. The binary BEC has a disk-shaped 
component along the $z$ direction  coupled
to a cigar-shaped component in the $(x,y)$ plane. 
For smaller nonlinearities of the 3D binary GPE,
we establish 
good agreement between the results for density and chemical potential 
of the effective BNLSE and the 
full 3D binary GPE. 
For larger nonlinearities
better result was obtained with the BNPSE. 
We also studied small oscillations of the binary 1D-2D BEC initiated by 
a sudden change of the radial  angular 
frequency of the quasi-2D disk-shaped BEC
using the BNPSE, BNLSE and the 
full 3D equations. 
Again the results obtained with the  BNPSE were in better agreement 
with those of the BNLSE.
Finally, a summary of 
our investigation is presented in Sec \ref{V}.  

\section{Theoretical Formulation for the BNLSE}

\label{II}

We shall now explicitly demonstrate the reduction procedure in three 
different types of binary mixtures:
1D-3D, 1D-2D, and 2D-3D. 
We shall consider a binary BEC where the trap anisotropy of 
the two components are distinct. When the component 2 has a fully 
asymmetric trap, the component 1, due to its specific 
trap symmetry, will be assumed to  have a 
quasi-1D or quasi-2D configuration. The quasi-1D shape emerges when the 
traps in the two transverse directions are much stronger than that in the 
linear direction. The quasi-2D shape emerges when the trap in the direction 
perpendicular to the 2D plane is much stronger than those in the 2D plane. 
In this fashion one could have an effective 
1D-3D, and 2D-3D mixed configuration for the traps acting on the binary 
BEC described by   effective 1D-3D and  2D-3D BNLSEs.  
There remains one distinct case where one of the components have a 
quasi-1D shape and the other component a quasi-2D shape due to specific 
trap symmetry of the two components. In that case the binary mixture is 
described by a  effective 1D-2D BNLSE. 
We discuss these possibilities in the following and illustrate the 
corresponding reduction schemes.

The action of a system of two BECs is given by \cite{emergent}
\beq 
A = \int ({\cal L}_1+{\cal L}_2-g_{12}|\psi_1|^2
|\psi_2|^2) \ d^3{\bf r}\ dt \; , 
\eeq
where the Lagrangian density is 
\beqa 
{\cal L}_j &=& \Big\{ i{\hbar\over 2} 
(\psi_j^* \partial_t\psi_j 
- \psi_j \partial_t\psi_j^* ) 
- {\hbar^2\over 2m_j} |\nabla_{\bf r} \psi_j|^2 
\nonumber
\\
&-& U_j({\bf r}) |\psi_j|^2  - {g_j\over 2} |\psi_j|^4 \Big\},  
\eeqa
with $U_j({\bf r})$ the external potential acting 
on the $j$-th BEC component ($j=1,2$), which is described by the macroscopic 
wave function $\psi_j({\bf r},t)$. Here $\partial_t$ denotes
time derivative, 
$g_j=4\pi \hbar^2a_j/m_j$, $g_{12}=2\pi \hbar^2 a_{12}/m_R$, and 
$\int |\psi_j({\bf r},t)|^2=N_j$, where $a_j$ is 
the intraspecies scattering length 
for species $j$ and $a_{12}$ is the interspecies scattering length, $m_j$ 
is the mass of species $j$,
and 
$m_R\equiv m_1m_2/(m_1+m_2)$ is the reduced mass. 

By extremizing the action $A$ with respect to $\psi_1^*({\bf r},t)$ and 
$\psi_2^*({\bf r},t)$ we obtain the following binary GPE \cite{emergent}
\beqa\label{3}
i\hbar \partial_t \psi_1 = 
\Big[ -{\hbar^2\over 2m_1}\nabla^2_{\bf r}  
+ U_1({\bf r}) + g_1 |\psi_1|^2 
+ g_{12} |\psi_2|^2 \Big] \psi_1 \; , 
\\
 \label{4}
i\hbar \partial_t \psi_2 = 
\Big[ -{\hbar^2\over 2m_2} \nabla^2 _{\bf r} 
+ U_2({\bf r}) + g_2 |\psi_2|^2 
+ g_{12} |\psi_1|^2 \Big] \psi_2 \; .  
\eeqa 
These two coupled partial differential equations are both 
in 3+1 (space-time) dimensions. In the general case, 
these can be solved numerically 
in a routine fashion \cite{murug,murug2}. 
However, in many situations of interest, when 
one or both of the BEC components  have extreme distinct 
cigar and disk shapes 
and when we are interested in density and dynamics in axial and radial 
directions, respectively, the following reduction procedures are of particular 
use. 

\subsection{1D-3D binary BEC} 

Let us suppose that the confining potential of the first BEC is  
\beq 
U_1({\bf r}) = {1\over 2} m_1 (\omega_{1x}^2 x^2 
+\omega_{1y}^2 y^2) + V_1(z), 
\label{pot1}
\eeq
where $V_1(z)$ is a generic potential in $z$ direction, and $\omega_{1x}$ 
and $\omega_{1y}$
are the angular frequencies of the harmonic trap along transverse $x$ and $y$ 
directions.  
The transverse harmonic potentials are assumed to be  so strong that the first 
BEC is quasi-1D, i.e. in an elongated 
cigar-shaped configuration.  

We can then use the variational wave function \cite{sala1} 
\beq 
\psi_1({\bf r},t) = 
{e^{-x^2/(2a_{1x}^2)}\over (\pi^{1/2} a_{1x})^{1/2} }
{e^{-y^2/(2a_{1y}^2)}\over (\pi^{1/2} a_{1y})^{1/2} } 
\ f_1(z,t) ,
\label{var1}
\eeq
for the first BEC, where $a_{1x}=\sqrt{\hbar/(m_1\omega_{1x})}$ and 
$a_{1y}=\sqrt{\hbar/(m_1\omega_{1y})}$ are the characteristic harmonic 
lengths of the oscillators in $x$ and $y$ directions, respectively. 
Consequently,  the action of the system becomes 
\beq 
A = \int L_1 \ dz dt + \int {\cal L}_2 \ d^3{\bf r} \ dt 
+ \int {\cal L}_{12} \ d^3{\bf r} \ dt \; , 
\eeq
where 
\beqa 
L_1 &=& i{\hbar\over 2} 
(f_1^* \partial_tf_1 
- f_1 \partial_tf_1^* ) 
- {\hbar^2\over 2m_1} |\partial_z f_1|^2  
\nonumber
\\
&-& V_1(z) |f_1|^2 -{\cal E}_1|f_1|^2 - {g_1\over 4\pi a_{1x} a_{1y}} |f_1|^4 ,
\label{L1}\\
{\cal L}_{12} &=& - {g_{12}\over \pi a_{1x } a_{1y } } 
e^{-(x^2/a_{1x}^2+y^2/a_{1y}^2)} |f_1|^2 |\psi_2|^2 \; ,\\
{\cal E}_1 &=& \frac{\hbar^2}{4m_1a_{1x}^2}+
\frac{m_1\omega_{1x}^2a_{1x}^2}{4} +
\frac{\hbar^2}{4m_1a_{1y}^2}+
\frac{m_1\omega_{1y}^2a_{1y}^2}{4}\; . \nonumber \\
\eeqa
We now extremize the action $A$ with respect to $f_1^*(z,t)$ and 
$\psi_2^*({\bf r},t)$, to obtain  the   effective BNLSE 
for 
$f_1(z,t)$ and $\psi_2({\bf r},t)$: 
\beqa 
i\hbar \partial_t f_1 &=& 
\Big[ -{\hbar^2\over 2m_1}\partial_z^2 
+ V_1(z) +{\cal E}_1 +{g_1\over 2\pi a_{1x}a_{1y}} |f_1|^2 
\nonumber 
\\
&+& {g_{12} \over \pi a_{1x} a_{1y}} 
\int e^{-(x^2/a_{1x}^2+y^2/a_{1y}^2)} 
|\psi_2|^2 dx dy \Big] f_1 \; , \nonumber \\  
\label{eqg1}\\
i\hbar \partial_t \psi_2 &=& 
\Big[ -{\hbar^2\over 2m_2} \nabla^2_{\bf r}  
+ U_2({\bf r}) + g_2 |\psi_2|^2 
\nonumber 
\\
&+& {g_{12}\over  \pi a_{1x } a_{1y}} 
e^{-(x^2/a_{1x}^2+y^2/a_{1y}^2)} 
|f_1|^2 \Big] \psi_2 \; .  
\label{eqg2}
\eeqa 
Equations (\ref{eqg1}) and (\ref{eqg2}) are simplified 
if we impose cylindrical  symmetry 
in the harmonic potential acting on the first BEC, i.e.  
$\omega_{1x}=\omega_{1y}=\omega_{1\rho}$, 
and also in the generic potential acting on the second 
BEC, i.e. $U_2({\bf r})=U_2(\rho,z)$ 
with $\rho=(x^2+y^2)^{1/2}$. In this way we can choose 
$\psi_2({\bf r},t)=\psi_2(\rho,z,t)$ and find
the following   effective BNLSE  for $f_1(z,t)$ 
and $\psi_2(\rho,z,t)$
\beqa \label{eq12}
i\hbar \partial_t f_1 &=& 
\Big[ -{\hbar^2\over 2m_1}\partial_z^2 
+ V_1(z)+{\cal E}_1 + {g_1\over 2\pi a_{1\rho}^2}  |f_1|^2 
\nonumber 
\\
&+& {2 g_{12} \over a_{1\rho}^2} 
\int_0^{\infty} e^{-\rho^2/a_{1\rho}^2} 
|\psi_2|^2 \rho d\rho \Big] f_1 \; , 
\\ \label{eq13}
i\hbar \partial_t \psi_2 &=& 
\Big[ -{\hbar^2\over 2m_2} 
\nabla_{\rho z}^2 
+ U_2(\rho,z) + g_2 |\psi_2|^2 
\nonumber 
\\
&+& {g_{12}\over  \pi a_{1\rho}^2} 
e^{-\rho^2/a_{1\rho}^2} |f_1|^2 \Big] \psi_2 \; , 
\eeqa 
where $a_{1\rho}\equiv a_{1x}\equiv a_{1y}
=\sqrt{\hbar/(m_1\omega_{1\rho})}$. In the coupled set of Eqs. (\ref{eq12})
and (\ref{eq13}), the first is integro-differential 
in 1+1 dimensions and the second is differential in 3+1 dimensions (but in 
2+1 variables). 

\subsection{1D-2D binary BEC} 

{\bf Parallel BECs}: 
Let us suppose that the confining potential of the first BEC 
is still given by Eq. (\ref{pot1}), while the confining potential 
of the second BEC is  
\beq 
U_2({\bf r}) = {1\over 2} m_2\omega_{2x}^2 x^2 +  V_2(y,z) \; , 
\label{pot2} 
\eeq
where $V_2(y,z)$ is a generic potential acting in the $(y,z)$ plane. 
In addition we impose that  
the harmonic potential of angular frequency $\omega_{2x}$ along the $x$ axis 
is so strong that the second BEC is quasi-2D, i.e. a very flat 
two-dimensional configuration in the $(y,z)$ plane.  Consequently, we have 
binary 1D-2D BEC where the first BEC is cigar-shaped along the $z$ axis 
and the second disk-shaped in the $(y,z)$ plane. 

We can then use the variational wave function of Eq. (\ref{var1}) 
for the first BEC, while for the second BEC we adopt the following 
variational wave function \cite{sala1}
\beq 
\psi_2({\bf r},t) =  
{e^{-x^2/(2a_{2x}^2)}\over (\pi^{1/2} a_{2x})^{1/2} } 
\ \phi_2(y,z,t) \; , 
\label{var2} 
\eeq
where $a_{2x}=\sqrt{\hbar/(m_2\omega_{2x})}$ is the characteristic harmonic 
length of the oscillator in $x$ direction
acting on the second BEC. 
In this way the action of the system becomes 
\beq 
A = \int L_1 \ dz dt + \int {\tilde {\cal L}}_2 \ dy dz dt 
+ \int {\tilde {\cal L}}_{12} \ dy dz dt \; , 
\eeq
where $L_1$ is still given by Eq. (\ref{L1}), while 
\beqa 
{\tilde {\cal L}}_2 &=& i{\hbar\over 2} 
(\phi_2^* \partial_t\phi_2 
- \phi_2 \partial_t\phi_2^* ) 
- {\hbar^2\over 2m_2} |\nabla_{yz} \phi_2|^2 
-\tilde{ \cal E}_2 |\phi_2|^2
\nonumber
\\
&-& V_2(y,z) |\phi_2|^2  
- {g_2\over 2 (2\pi)^{1/2} a_{2x}} |\phi_2|^4, \\
\tilde { \cal E}_2&=&\frac{\hbar^2}{4m_2a_{2x}^2}
+\frac{m_2\omega_{2x}^2a_{2x}^2}{4}    \\
{\tilde {\cal L}}_{12} &=& 
- {g_{12}\over \pi a_{1y } \sqrt{a_{1x }^2+a_{2x}^2} } 
|f_1|^2 |\phi_2|^2 e^{-y^2/a_{1y}^2} \; . 
\eeqa
We now extremize the action $A$ and find the following 
effective BNLSE for 
$f_1(z,t)$ and $\phi_2(y,z,t)$:
\beqa 
i\hbar \partial_t f_1 = 
\Big[ -{\hbar^2\over 2m_1}\partial_z^2 
+ V_1(z)+{\cal E}_1 + {g_1\over 2\pi a_{1x}a_{1y}} |f_1|^2
\nonumber 
\\
+ 
{g_{12}\over \pi a_{1y }\sqrt{a_{1x }^2+a_{2x}^2}} 
\int_{-\infty}^{\infty} 
e^{-y^2/a_{1y}^2} |\phi_2|^2 dy \Big] f_1 \; , 
\label{eq1}\\
i\hbar \partial_t \phi_2 = 
\Big[ -{\hbar^2\over 2m_2} \nabla_{yz}^2 + V_2(y,z)+\tilde{ \cal E}_2 
+{g_2\over \sqrt{2\pi} a_{2x}} |\phi_2|^2 
\nonumber 
\\
+ {g_{12}e^{-y^2/a_{1y}^2}
\over \pi a_{1y }\sqrt{a_{1x }^2+a_{2x}^2} } 
|f_1|^2 \Big] \phi_2 \; .  
\label{eq2}
\eeqa  
These are two coupled equations, where the first is integro-differential 
in 1+1 dimensions and the second is differential in 2+1 dimensions.
Equations  (\ref{eq1}) and (\ref{eq2}) 
describe the  binary system where the axis of the cigar-shaped BEC along
$z$ direction 
is in the $(y,z)$  plane of the two-dimensional BEC. 

{\bf Perpendicular BECs}: 
Another interesting possibility is a binary system where 
the cigar-shaped BEC is perpendicular to the plane of the two-dimensional 
BEC. This kind of configuration can be obtained by considering the 
following trapping potential acting on the second BEC 
\beq 
U_2({\bf r}) = W_2(x,y) + {1\over 2} m_2\omega_{2z}^2 z^2  \; , 
\label{pot-mio}
\eeq
where $W_2(x,y)$ is a generic potential in $(x,y)$ plane,  
while the harmonic potential of angular frequency $\omega_{2z}$ 
along the $z$ axis is so strong that the second BEC is a quasi-2D 
in the $(x,y)$ plane. 
For the first BEC we still use the potential of Eq. (\ref{pot1}) and 
the variational wave function of Eq. (\ref{var1}), but 
for the second BEC we adopt the variational wave function 
\beq 
\psi_2({\bf r},t) = h_2(x,y,t) \ 
{e^{-z^2/(2a_{2z}^2)}\over (\pi^{1/2} a_{2z})^{1/2} } \; , 
\label{var-mio}
\eeq
where $a_{2z}=\sqrt{\hbar/(m_2\omega_{2z})}$ is the characteristic harmonic 
length of the potential  acting on the second BEC along $z$ direction. 
From Eqs. (\ref{eqg1}) and (\ref{eqg2}), using Eq. (\ref{var-mio}),  
it is straightforward 
to derive the effective BNLSE for $f_1(z,t)$ and $h_2(x,y,t)$ \cite{sala1}: 
\beqa 
&&i\hbar \partial_t f_1 = 
\Big[ -{\hbar^2\over 2m_1}\partial_z^2 
+ V_1(z) +{\cal E}_1 +{g_1\over 2\pi a_{1x}a_{1y}} 
|f_1|^2  
\nonumber 
\\
&&+{g_{12}  e^{-z^2/a_{2z}^2} 
\over \pi^{3/2} a_{1x} a_{1y} a_{2z}}
\int_{-\infty}^{\infty} 
e^{-(x^2/a_{1x}^2+y^2/a_{1y}^2)} 
|h_2|^2 dx dy \Big] f_1 \; ,
\label{eg1}\nonumber \\
\\
&&i\hbar \partial_t h_2 = 
\Big[ -{\hbar^2\over 2m_2} \nabla_{xy}^2 
+ W_2(x,y) + {g_2\over (2\pi)^{1/2} a_{2z}} |h_2|^2 
\nonumber 
\\
&&+E_2+g_{12} {e^{-(x^2/a_{1x}^2+y^2/a_{1y}^2)} 
\over  \pi^{3/2} a_{1x } a_{1y}a_{2z}} 
\int_{-\infty}^{\infty} 
e^{-z^2/a_{2z}^2} |f_1|^2 dz \Big] h_2 \;\label{eg2}
 ,\nonumber \\
\\
&& {  E}_2=\frac{\hbar^2}{4m_2a_{2z}^2}
+\frac{m_2\omega_{2z}^2a_{2z}^2}{4}   \; .
\eeqa 
Equations  (\ref{eg1}) and (\ref{eg2}) are simplified 
if we impose cylindrical symmetry 
in the harmonic potential acting on the first BEC, i.e.  
$\omega_{1x}=\omega_{1y}=\omega_{1\rho}$, 
and also in the generic potential acting the second BEC, 
i.e. $W_2(x,y)=W_2(\rho)$ 
with $\rho=(x^2+y^2)^{1/2}$. In this way we can choose 
$h_2(x,y,t)=h_2(\rho,t)$ and find the following   effective 
BNLSE for $f_1(z,t)$ and $ h_2(\rho,t)$:
\beqa 
i\hbar \partial_t f_1 &=& 
\Big[ -{\hbar^2\over 2m_1}\partial_z^2 
+ V_1(z)+{\cal E}_1  + {g_1\over 2\pi a_{1\rho}^2} |f_1|^2+{\cal E}_1 
\nonumber 
\\
&+& {2 g_{12}  e^{-z^2/a_{2z}^2} 
\over \pi^{1/2} a_{1\rho}^2 a_{2z}}
\int_0^{\infty} e^{-\rho^2/a_{1\rho}^2} 
|h_2|^2 \rho d\rho \Big] f_1 \; ,\label{25} \\
\label{26}
i\hbar \partial_t h_2 &= &
\Big[ -{\hbar^2\over 2m_2} \nabla_{\rho}^2 
+ W_2(\rho)+E_2 + {g_2\over (2\pi)^{1/2} a_{2z}} |h_2|^2 
\nonumber 
\\
&+& g_{12} {e^{-\rho^2/a_{1\rho}^2} 
\over  \pi^{3/2} a_{1\rho}^2
{a_{2z}}  
} \int_{-\infty}^{\infty} e^{-z^2/a_{2z}^2} 
|f_1|^2 dz \Big] h_2 \; .  
\eeqa 
These are two coupled equations, where the first equation 
 is integro-differential 
in 1+1 dimensions and the second is integro-differential in 2+1 dimensions
(but in 1+1 variables).

\subsection{3D-2D binary BEC} 

Let us suppose now that the first BEC is fully 3D 
and it is under the action of the generic potential $U_1({\bf r})$, 
while the second BEC is trapped by the potential (\ref{pot-mio}).  
Again we assume that the harmonic potential of angular frequency $\omega_{2z}$ 
along the $z$ axis is so strong that the second BEC is a quasi-2D 
 in the $(x,y)$ plane. The first BEC is described by the 
generic wave function $\psi_1({\bf r},t)$ while 
the second BEC is modeled by the variational wave function 
of Eq. (\ref{var-mio}). Following the same procedure above,   
the wave functions $\psi_1({\bf r},t)$ and $h_2(x,y,t)$ are found to satisfy 
the following  effective BNLSE 
\beqa
i\hbar \partial_t \psi_1 &= &
\Big[ -{\hbar^2\over 2m_1}\nabla^2_{\bf r}  
+ U_1({\bf r}) + g_1 |\psi_1|^2 
\nonumber
\\
&+& {g_{12}\over \pi^{1/2} a_{2z}} e^{-z^2/a_{2z}^2} 
|h_2|^2 \Big] \psi_1 \; , \\
i\hbar \partial_t h_2 &=& 
\Big[ -{\hbar^2\over 2m_2} \nabla_{xy}^2 
+ W_2(x,y) + {g_2\over (2\pi)^{1/2}a_{2z}} |h_2|^2 
\nonumber 
\\
&+& E_2+{g_{12}\over \pi^{1/2}a_{2z}} 
\int_{-\infty}^{\infty} e^{-z^2/a_{2z}^2} 
|\psi_1|^2 dz \Big] h_2 \; .  
\eeqa 
As in the previous section, these equations are simplified 
if the confining potentials have cylindrical symmetry: 
$U_1({\bf r})=U_1(\rho,z)$ and 
$W_2(x,y)=W_2(\rho)$.  In this way we can choose 
$\psi_1({\bf r},t)=\psi_1(\rho,z,t)$ and 
$h_2(x,y,t)=h_2(\rho,t)$ and the effective  BNLSE become 
\beqa
i\hbar \partial_t \psi_1 &=& 
\Big[ -{\hbar^2\over 2m_1}\left( \nabla_{\rho}^2 + \partial_z^2 \right) 
+ U_1(\rho,z) + g_1 |\psi_1|^2 
\nonumber
\\
&+& {g_{12}\over \pi^{1/2} a_z} e^{-z^2/a_{2z}^2} 
|h_2|^2 \Big] \psi_1 \; , \\
i\hbar \partial_t h_2 &= &
\Big[ -{\hbar^2\over 2m_2} { \nabla_{\rho}^2 } 
+ W_2(\rho)+E_2 + {g_2\over (2\pi)^{1/2}a_{2z}} |h_2|^2 
\nonumber 
\\
&+& {g_{12}\over \pi^{1/2}a_{2z}} 
\int_{-\infty}^{\infty} e^{-z^2/a_{2z}^2} 
|\psi_1|^2 dz \Big] h_2 \; .  
\eeqa 
These are two coupled equations, where the first is differential 
in 3+1 dimensions (but in 2+1 variables) 
and the second is integro-differential in 2+1 dimensions 
(but in 1+1 variables).

\section{Dimensional reduction beyond the effective BNLSE: the BNPSE}

\label{III}

In Sec. \ref{II} we considered dimensional reduction of the GPE
 for binary BEC in the form of an effective BNLSE in reduced 
dimensions. Such equations are simpler than the 3D binary GPE 
 and  they could be 
good approximation to the original equations. It is possible to 
improve on the effective BNLSE presented in Sec. \ref{II} through 
a binary nonpolynomial Schr\"odinger equation (BNPSE). 
This improvement is obtained 
by considering a more flexible variational wave function in place of 
Eqs. (\ref{var1}), (\ref{var2}), and (\ref{var-mio}), where in the 
transverse direction the wave function is taken as the ground state in the 
respective harmonic potential(s) with Gaussian form. In the following we 
maintain the Gaussian form but with a flexible width to obtain the required 
BNPSEs in two cases: (i) The axially symmetric 1D-3D case described by 
Eqs. (\ref{eq12}) and (\ref{eq13}) and (ii) The 
axially symmetric perpendicular 1D-2D case described by Eqs. (\ref{25})
and (\ref{26}).   
The other cases presented in Sec. \ref{II} can be treated in a similar 
and straight-forward fashion and the respective BNPSE obtained. 

\subsection{1D-3D with cylindrical symmetry}

Here we would like to introduce the next-order correction to Eqs. 
(\ref{eq12})  and (\ref{eq13}). In this case
the confining potential (\ref{pot1}) of the first BEC can be written as 
\beq
U_1({\bf r}) = {1\over 2} m_1 \omega_{1\rho}^2 \rho^2 + V_1(z).
\label{eq34}
\eeq

We can then use the variational wave function
\beq
\psi_1({\bf r},t) =
{e^{-\rho^2/(2\sigma^2(z,t))}
\over \pi^{1/2} \sigma(z,t) } \ f_1(z,t),
\label{eq35}
\eeq
in place of Eq. (\ref{var1}), 
for the first BEC, where $\sigma\equiv \sigma(z,t)$ is its width,
which can be quite different from the characteristic harmonic
length $a_{\rho}=\sqrt{\hbar/(m_1\omega_{1\rho})}$
of the transverse confinement. 
We assume, consistent with the philosophy 
of the reduction scheme,  that in Eq. (\ref{eq35}) the $z$ and $t$ derivatives 
act only on $f_1$.
Then the action of the system becomes
\beq
A = \int \hat L_1 \ dz dt + \int \hat {\cal L}_2 \ d^3{\bf r} \ dt
+ \int \hat {\cal L}_{12} \ d^3{\bf r} \ dt \; ,
\eeq
where
\beqa\label{eq37}
\hat L_1 &=& i{\hbar\over 2}
(f_1^* \partial_tf_1
- f_1 \partial_tf_1^* )
- {\hbar^2\over 2m_1} |\partial_z f_1|^2- V_1(z) |f_1|^2
\nonumber
\\
&-& \left({\hbar^2\over 2m_1\sigma^2} +
{m_1\omega_{1\rho}^2\sigma^2\over 2}\right){|f_1|^2}
- {g_1\over 4\pi \sigma^2} |f_1|^4,
\label{L1x}\\
\hat {\cal L}_2 &=& i{\hbar\over 2}
(\psi_2^* \partial_t\psi_2
- \psi_2 \partial_t\psi_2^* )
- {\hbar^2\over 2m_2} |\nabla \psi_2|^2
\nonumber
\\
&-& U_2(\rho,z) |\psi_2|^2  - \frac{g_2}{2} |\psi_2|^4,\\
\hat {\cal L}_{12} &=& - {g_{12}\over \pi \sigma^2 }
e^{-\rho^2/\sigma^2} |f_1|^2 |\psi_2|^2 \; ,
\eeqa
where  we have neglected, in the spirit of the reduction scheme, 
the spatial derivative of $\sigma$.

We now extremize the action $A$ with respect to $f_1^*(z,t)$,
$\psi_2^*(\rho,z,t)$ and $\sigma(z,t)$, finding the equations
for $f_1(z,t)$, $\psi_2(\rho,z,t)$ and $\sigma(z,t)$:
\beqa \label{eq40}
i\hbar \partial_t f_1 &=&
\Big[ -{\hbar^2\over 2m_1}\partial_z^2
+ V_1(z) + {g_1\over 2\pi \sigma^2}  |f_1|^2
\nonumber
\\  
&+&
{\left( \frac{\hbar^2}{2m_1\sigma^2}+\frac{m_1\omega_{1\rho}^2\sigma^2}
{2}                   \right)}
\nonumber
\\
&+& {2 g_{12} \over \sigma^2}
\int_0^{\infty} e^{-\rho^2/\sigma^2}
|\psi_2(\rho,z,t)|^2 \rho d\rho \Big] f_1 \; ,\\
i\hbar \partial_t \psi_2 &=&
\Big[ -{\hbar^2\over 2m_2} \nabla_{\rho z}^2
+ U_2(\rho,z) + g_2 |\psi_2|^2
\nonumber
\\
&+& {g_{12}\over  \pi \sigma^2}
e^{-\rho^2/\sigma^2} |f_1(z,t)|^2 \Big] \psi_2 \; ,  \label{eq41}  \\
m_1\omega_{1\rho}^2 \sigma^4 &=& {\hbar^2 \over m_1} +
{g_1\over 2\pi} |f_1|^{2} 
\nonumber
\\
&+&{4}g_{12} \int_0^{\infty}
d\rho \rho e^{-\rho^2/\sigma^2} {|\psi_2|^2}
\left( 1 {-} {\rho^2\over \sigma^2} \right) \; . \label{eq42}
\eeqa
Equations (\ref{eq40}) and  (\ref{eq41})
are the BNPSE, where the first is integro-differential
in 1+1 dimensions, the second is differential in 3+1 dimensions (but in 
2+1 variables). Equation (\ref{eq42})
is integro-algebraic and determines the width 
$\sigma$ as a function of $|f_1|^2$ and 
$|\psi_2|^2$. If we take $\sigma= a_{1\rho}$, the harmonic length in the 
transverse $\rho$ direction for the first component, in place of Eq. 
(\ref{eq42}), we get the previously obtained effective 1D-3D Eqs. (\ref{eq12})
and (\ref{eq13}).  
If we set the coupling term $g_{12}=0$ in Eq. (\ref{eq42}) 
we obtain the model previously obtained in 
Ref. \cite{sala1} for a cigar-shaped BEC.

\subsection{Perpendicular 1D-2D with cylindrical symmetry}

Here we would like to introduce the next-order correction to Eqs. 
(\ref{25}) and (\ref{26}). In this case the confining potential is given 
by Eq. (\ref{eq34}) and the variational wave function of the first BEC 
will be taken as (\ref{eq35}). The potential of the second BEC will be taken as
\begin{equation}\label{eq43}
U_2({\bf r})=W_2(\rho)+\frac{1}{2}m_2\omega_{2z}^2 z^2,
\end{equation}
and its variational wave function as 
\begin{equation}\label{eq44}
\psi_2({\bf r},t)=\frac{e^{-z^2/(2\eta^2(\rho,t))}}
{\pi^{1/4}\eta^{1/2}(\rho,t)}
h_2(\rho,t),
\end{equation}
where the width $\eta\equiv \eta(\rho,t)$ can be quite 
different from the characteristic harmonic length along the $z$ direction and 
the $\rho$ and $t$ derivatives are supposed to act only on $h_2$.  

The action of the system can now be written as 
\beq 
A = \int \hat L_1 \ dz dt +2 \pi \int \bar { L}_2 \ \rho d\rho \ dt
+ \int \bar {\cal L}_{12} \ d^3{\bf r} \ dt \; ,
\eeq
where $\hat L_1$ is given by Eq. (\ref{eq37}) and 
\beqa
\bar  L_2 &=& i{\hbar\over 2}
(h_2^* \partial_t h_2
- h_2 \partial_t h_2^* )
- {\hbar^2\over 2m_2} |\nabla_\rho h_2|^2
- W_2(\rho) |h_2|^2  
\nonumber \\
&-&\frac{1}{2}\left({\hbar^2\over 2m_2\eta^2} +
{m_2\omega_{2z}^2\eta^2\over 2}\right){|h_2|^2}
- \frac{g_2}{2\eta\sqrt{2\pi}} |h_2|^4,\\
\bar {\cal L}_{12}&=& -\frac{g_{12}}{\pi^{3/4}\eta\sigma^2}e^{-\rho^2/\sigma^2}
e^{-z^2/\eta^2}
|f_1|^2|h_2|^2.
\eeqa
Again we have neglected the space derivatives of the widths $\sigma$ 
and $\eta$.
We now extremize the action  $A$ with respect to $f_1^*$, $h_2^*$, $\sigma$ 
and $\eta$ to get the desired equations
\beqa \label{eq48}&&
i\hbar \partial_t f_1 =
\Big[ -{\hbar^2\over 2m_1}\partial_z^2
+ V_1(z) + {g_1\over 2\pi \sigma^2} |f_1|^2
\nonumber
\\  
&+&
{\left( \frac{\hbar^2}{2m_1\sigma^2}+\frac{m_1\omega_{1\rho}^2\sigma^2}
{2}                   \right)}
\nonumber
\\
&+& {2 g_{12}  e^{-z^2/\eta^2}
\over \pi^{1/2} \sigma^2 \eta}
\int_0^{\infty} e^{-\rho^2/\sigma^2} 
|h_2(\rho,t)|^2 \rho d\rho \Big] f_1 \; ,
\\ \label{eq49}&&
i\hbar \partial_t h_2 =
\Big[ -{\hbar^2\over 2m_2} \nabla_{\rho}^2
+ W_2(\rho) + {g_2\over (2\pi)^{1/2} \eta} |h_2|^2
\nonumber
\\  
&+&
\frac{1}{2}
{\left( \frac{\hbar^2}{2m_2\eta^2}+\frac{m_2\omega_{2z}^2\eta^2}
{2}                   \right)}
\nonumber
\\
&+& g_{12} {e^{-\rho^2/\sigma^2}
\over  \pi^{3/2} \sigma^2\eta
} \int_{-\infty}^{\infty} e^{-z^2/\eta^2}
|f_1(z,t)|^2 dz \Big] h_2 \; ,
\\  \label{eq50}&&
m_1\omega_{1\rho}^2 \sigma^4 = {\hbar^2 \over m_1} +
{g_1\over 2\pi} |f_1|^2 +
{4g_{12}} \pi^{1/4}\int_0^{\infty}
d\rho \rho \frac{1}{\eta}
\nonumber
\\
&\times&
e^{-\rho^2/\sigma^2} e^{-z^2/\eta^2} |h_2|^2
\left( 1 - {\rho^2\over \sigma^2} \right) \; .\\
\label{eq51}&&
m_2\omega_{2z}^2 \eta^4 = {\hbar^2 \over m_2} +
{g_2\eta \over 2\sqrt{2\pi}} |h_2|^2 
+\frac{g_{12}\eta}{\pi^{3/4}}\int_{-\infty}^{\infty}
dz \frac{1}{\sigma^2} 
\nonumber
\\
&\times& e^{-\rho^2/\sigma^2} e^{-z^2/\eta^2} |f_1|^2
\left( 1 - 2{z^2\over \eta^2} \right) \; .
\eeqa
Equations (\ref{eq48}) and  (\ref{eq49}) constitute the BNPSE, 
where the first one is integro-differential in 1+1 dimensions, the second 
is  integro-differential in 2+1 dimensions (but in 1+1 variables).
Equations (\ref{eq50}) and  (\ref{eq51}) 
integro-algebraic and determines the widths $\sigma$ and $\eta$ as 
functions of $|f_1|^2$ and $|h_2|^2$. If we take $\sigma=a_{1\rho}$ and 
$\eta=a_{2z}$, the harmonic lengths in the transverse directions for the first 
and the second components, respectively, in place of Eqs. (\ref{eq50})
and (\ref{eq51}), we get back Eqs. (\ref{25}) and (\ref{26}). 
If we set the coupling 
term $g_{12}=0$ in Eqs. (\ref{eq50}) and (\ref{eq51}),
Eqs. (\ref{eq48}) and (\ref{eq49}) reduce to
 previously studied models 
in Ref. \cite{sala1} for a cigar- and disk-shaped BECs, respectively.

\section{Numerical Results}

We shall present results for the coupled 1D-2D case when the 
cigar-shaped BEC is perpendicular to the disk-shaped BEC. The 
potentials $V_1(z) $ and $W_2(\rho)$ in Eqs. (\ref{25}) and (\ref{26})
are taken as \begin{equation} 
V_1(z)=\frac{1}{2}m_1\omega_{1z}^2z^2; \quad 
W_2(\rho)=\frac{1}{2}m_2\omega_{2\rho}^2 \rho^2. \end{equation}
The cigar-shaped BEC, labeled 1, lies along the $z$ 
axis and has trapping frequencies satisfying 
$\omega_{1\rho}/\omega_{1z}=10$. 
The disk-shaped BEC, labeled 2, 
lies in the $(x,y)$ plane and has trapping frequencies satisfying 
$\omega_{2z}/\omega_{2\rho}=10$.  

\begin{figure}
\begin{center}
\includegraphics[width=.7\linewidth]{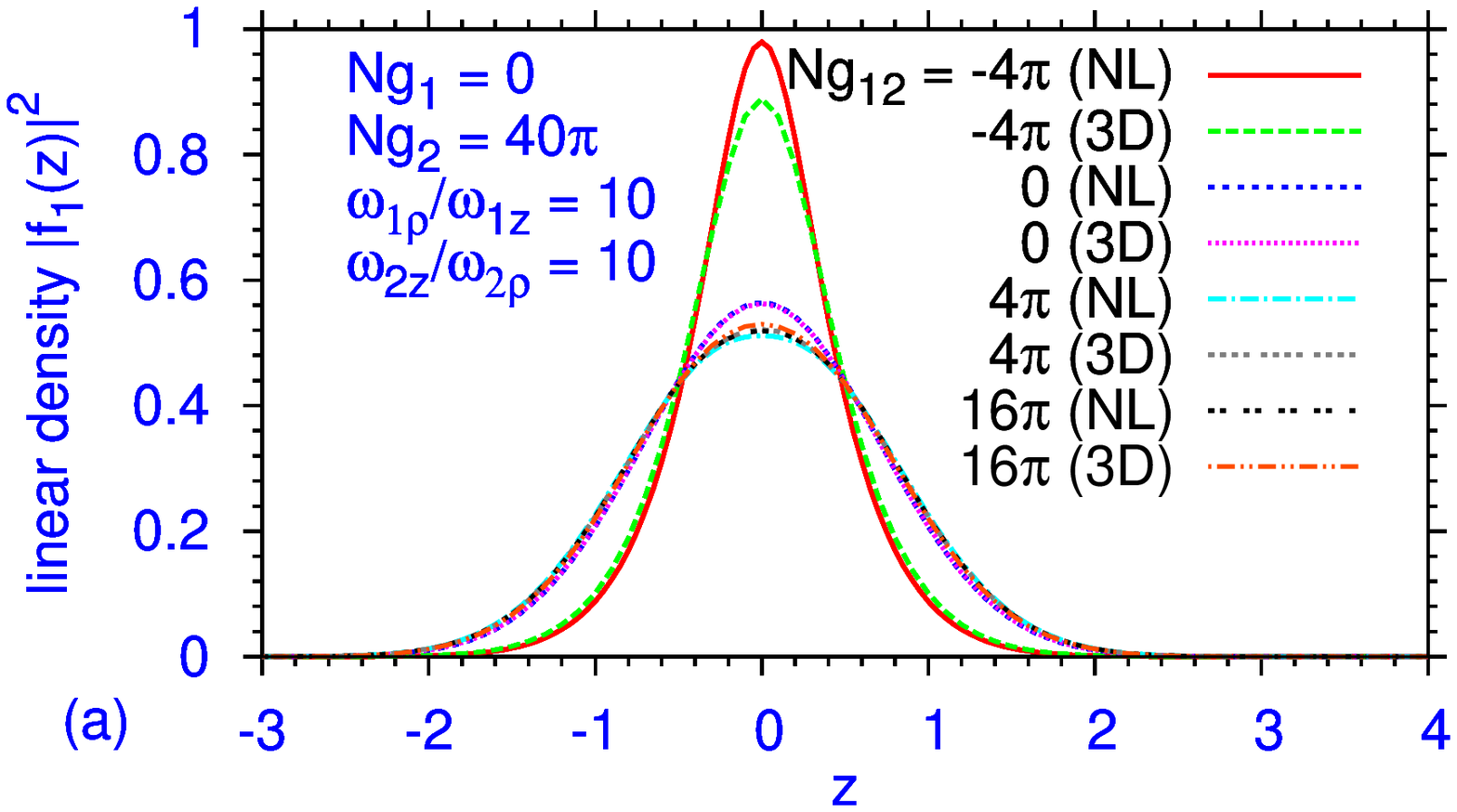}
\includegraphics[width=.7\linewidth]{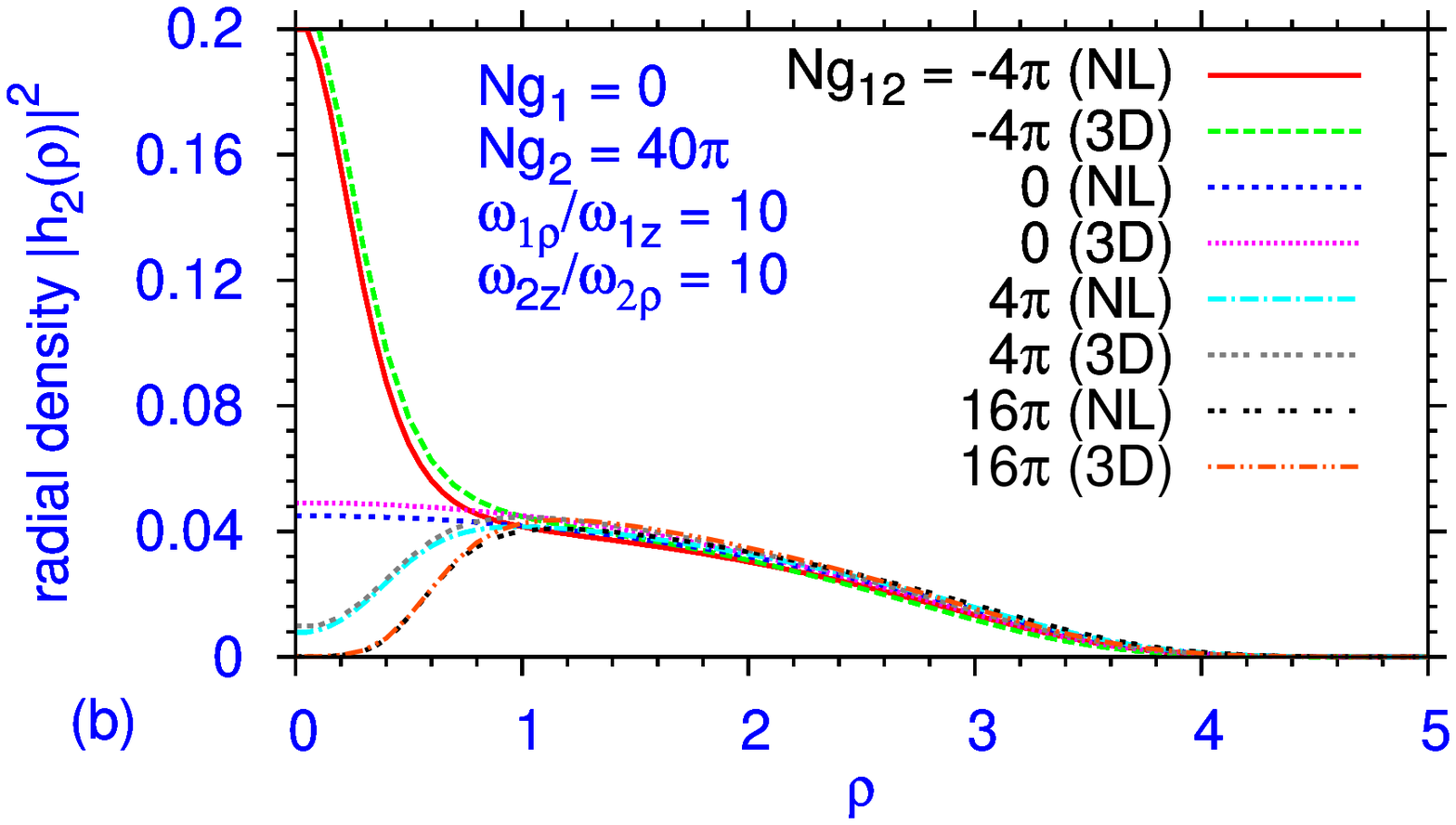}
\end{center}

\caption{(Color online) Plot of 
(a) linear and (b) radial  densities of a binary perpendicular 
cigar-disk mixture as obtained 
from the 3D GPE (\ref{3}) and (\ref{4}) (3D)
and the reduced 1D-2D BNLSE 
(\ref{25}) and  (\ref{26}) (NL)
in dimensionless units 
for $Ng_1=0$ and $Ng_2=40\pi$ for different 
$Ng_{12}$.   Parameters used: $m_1=m_2=\hbar=1$, $\omega_{1\rho}=
\omega_{2z}=10, \omega_{1z}=
\omega_{2\rho}=1,$ $a_{1z}=a_{2\rho}=1$, and $a_{1\rho}=a_{2z}=1/
\sqrt{10}$. Both densities are normalized to unity.
}
\label{fig1}
\end{figure}

To keep the algebra under control, we work in dimensionless units and 
set $m_1=m_2=\hbar=1$ and take $N_1=N_2=N$. 
For the cigar-shaped BEC we take $\omega_{1\rho}=10$ 
and   $\omega_{1z}=1$; for the disk-shaped BEC we take  $\omega_{2z}=10$,
and  $\omega_{2\rho}=1$. 
For the cigar-shaped BEC 1, the  harmonic oscillator lengths are 
$a_{1z} \equiv \sqrt{\hbar /(m\omega_{1z})}=1$ and $a_{1\rho}
\equiv  \sqrt{\hbar /(m\omega_{1\rho})}=1/\sqrt{10}$. 
For the disk-shaped BEC 2, 
the  harmonic oscillator lengths are 
$a_{2z} \equiv \sqrt{\hbar /(m\omega_{2z})}= 1/\sqrt{10}$ and $a_{2\rho}
\equiv  \sqrt{\hbar /(m\omega_{2\rho})}=1$.  The linear and radial densities 
are calculated from the solution of Eqs. (\ref{3}) and (\ref{4}) via
\begin{eqnarray}
|f_1(z)|^2= 2\pi\int_0^\infty \rho d\rho |\psi_1({\bf r})|^2,\\
|h_2(\rho)|^2= \int_{-\infty}^\infty  dz |\psi_1({\bf r})|^2,
\end{eqnarray}
to be compared with the solutions of Eqs. (\ref{25}) and (\ref{26}).

\begin{figure}
\begin{center} 
\includegraphics[width=.7\linewidth]{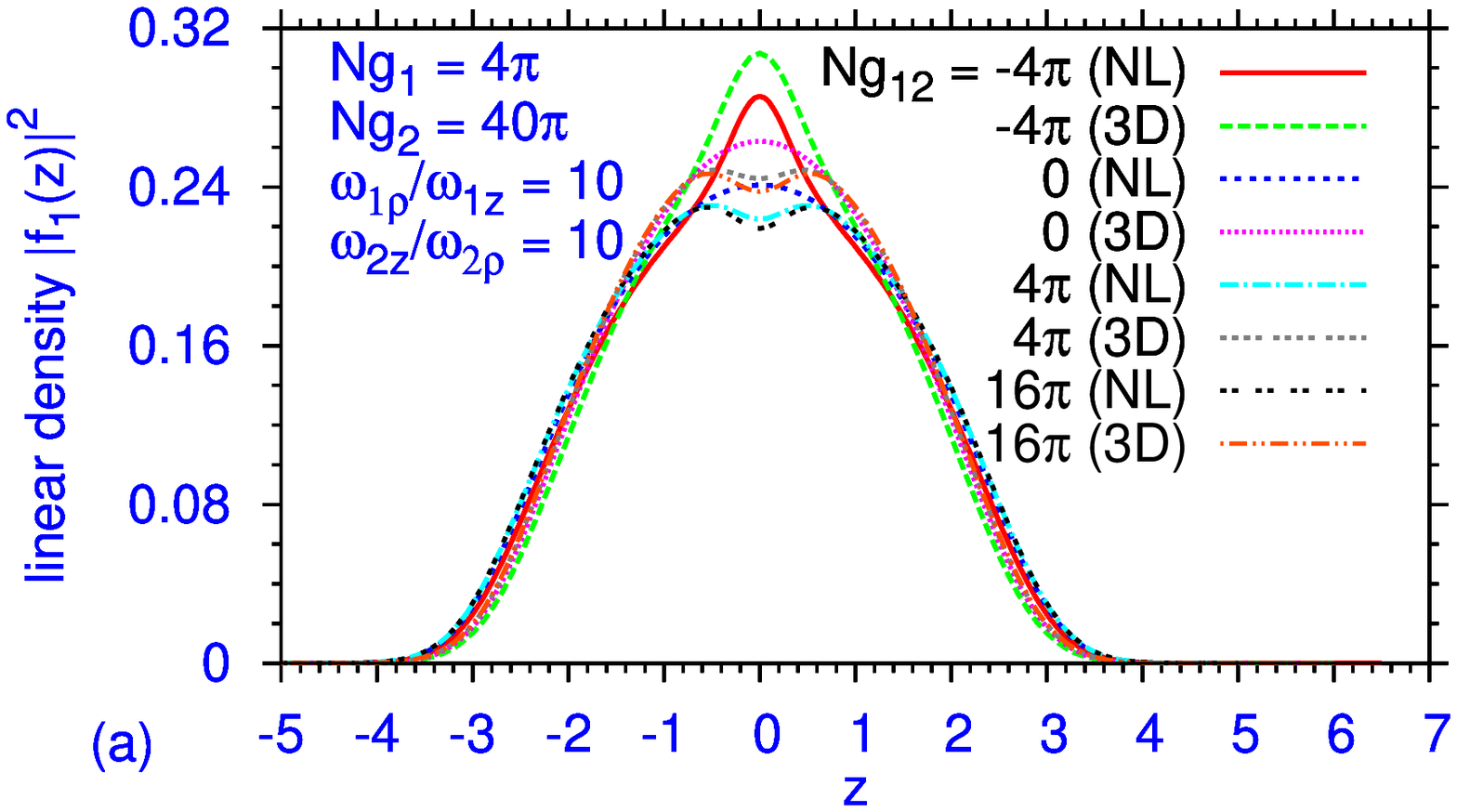} 
\includegraphics[width=.7\linewidth]{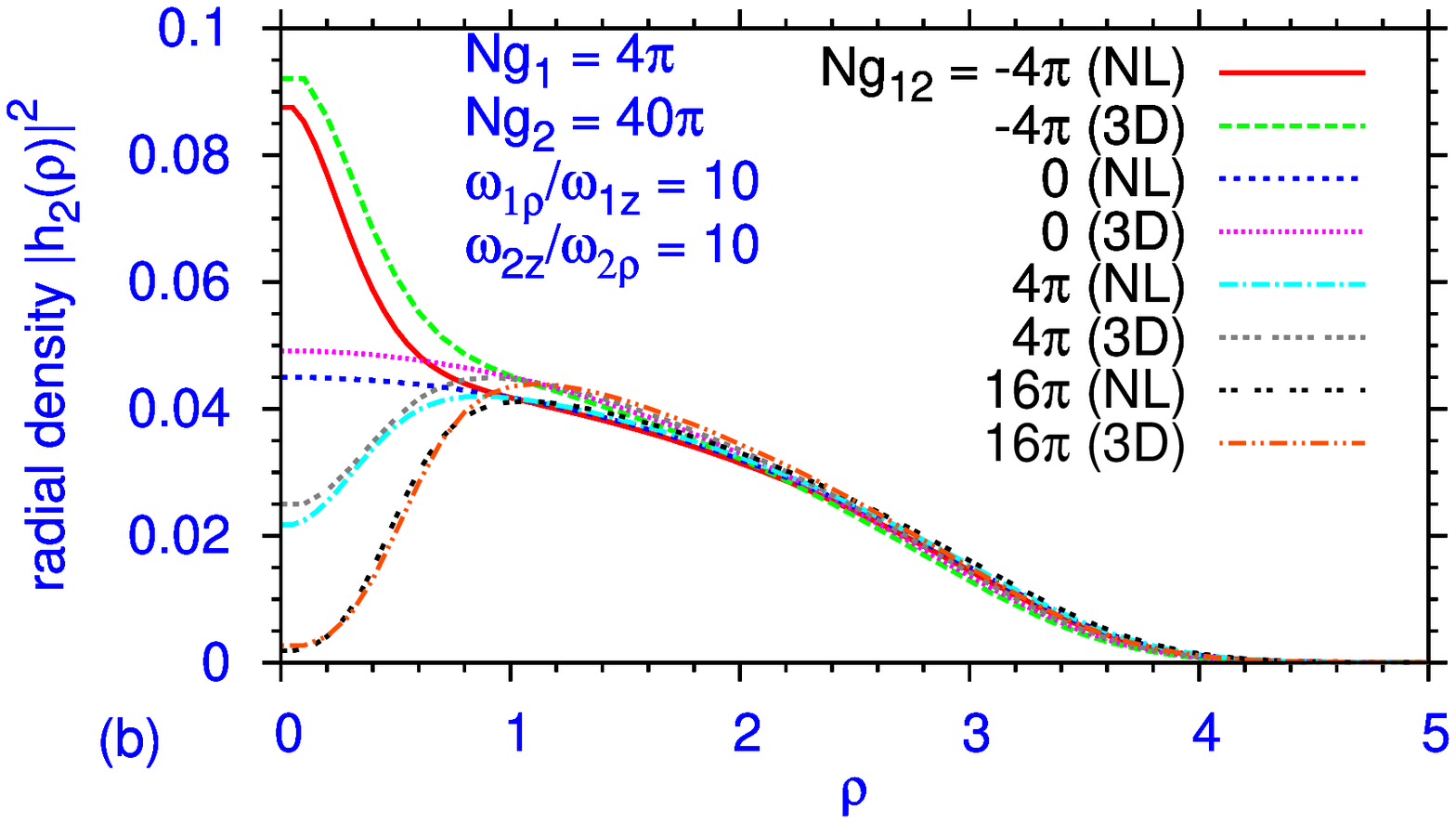} \end{center}
\caption{(Color online) The same as in Fig. \ref{fig1} for $Ng_1
=4\pi$ and $Ng_2=40\pi$.
}
\label{fig2}
\end{figure}

\begin{figure}
\begin{center}
\includegraphics[width=.7\linewidth]{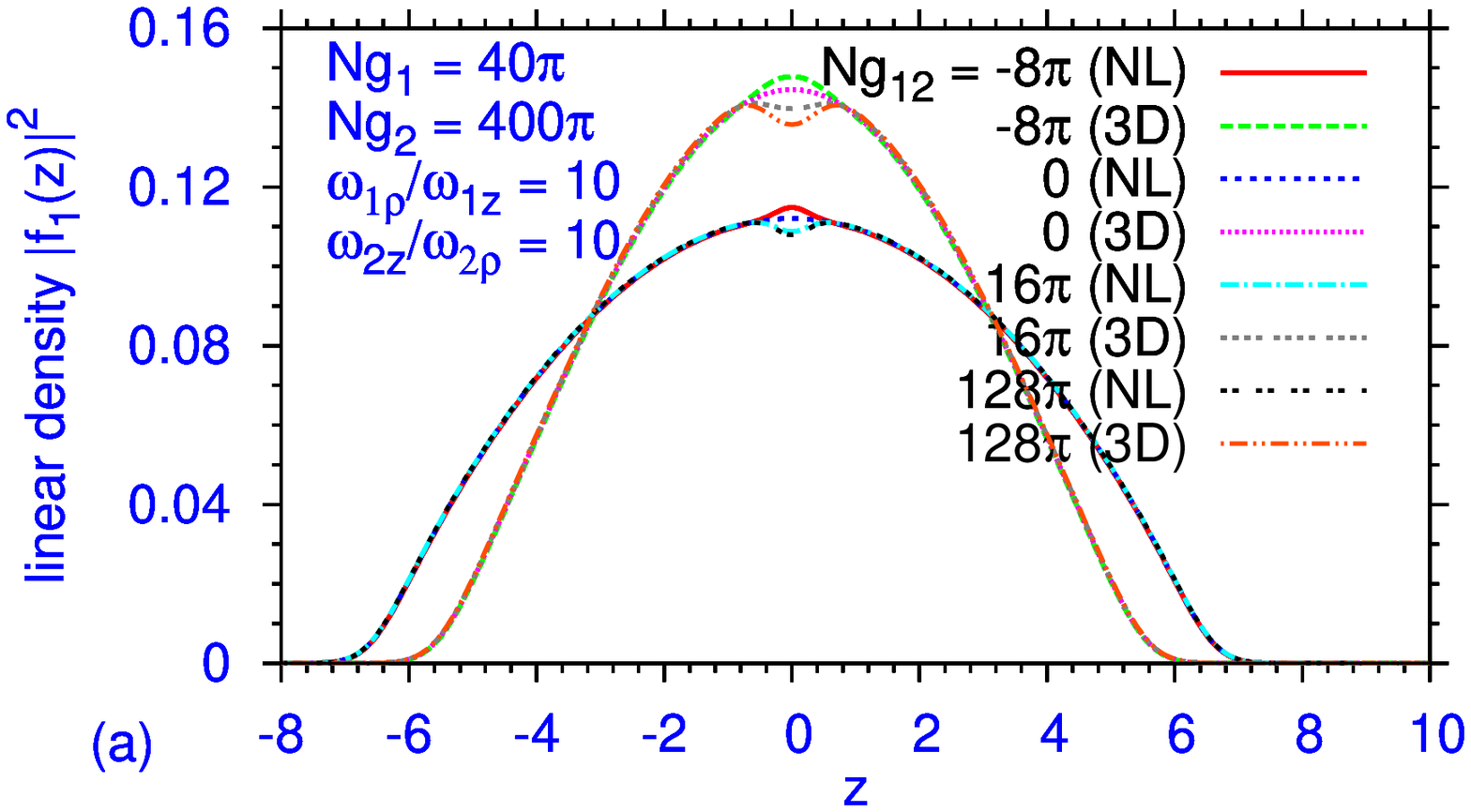}
\includegraphics[width=.7\linewidth]{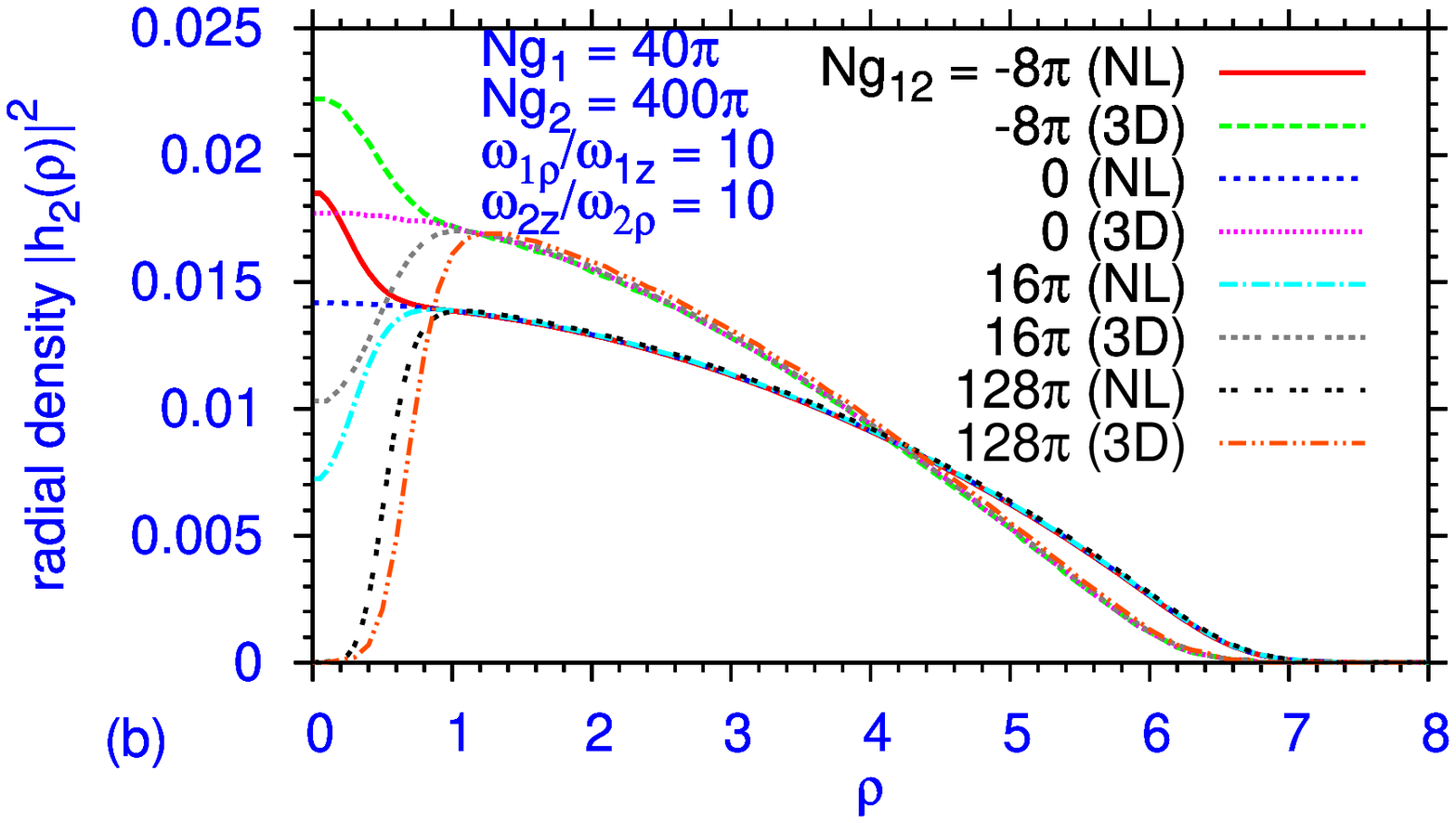}
\end{center}

\caption{(Color online) The same as in Fig. \ref{fig1} for $Ng_1
=4\pi$ and $Ng_2=400\pi$.
}

\label{fig3}
\end{figure}

To solve the coupled GPEs in three and reduced dimensions we 
employ the split-step Crank-Nicolson discretization technique 
\cite{murug} with real and imaginary time propagation. We essentially 
use the Fortran programs provided in Ref. \cite{murug}. The convergence 
of the numerical result was tested by varying the space (typically $\sim$ 0.05)
and time (typically $\sim$ 0.001)
steps 
and the total number of spatial and temporal discretization points. 

\begin{figure}
\begin{center}
\includegraphics[width=.7\linewidth]{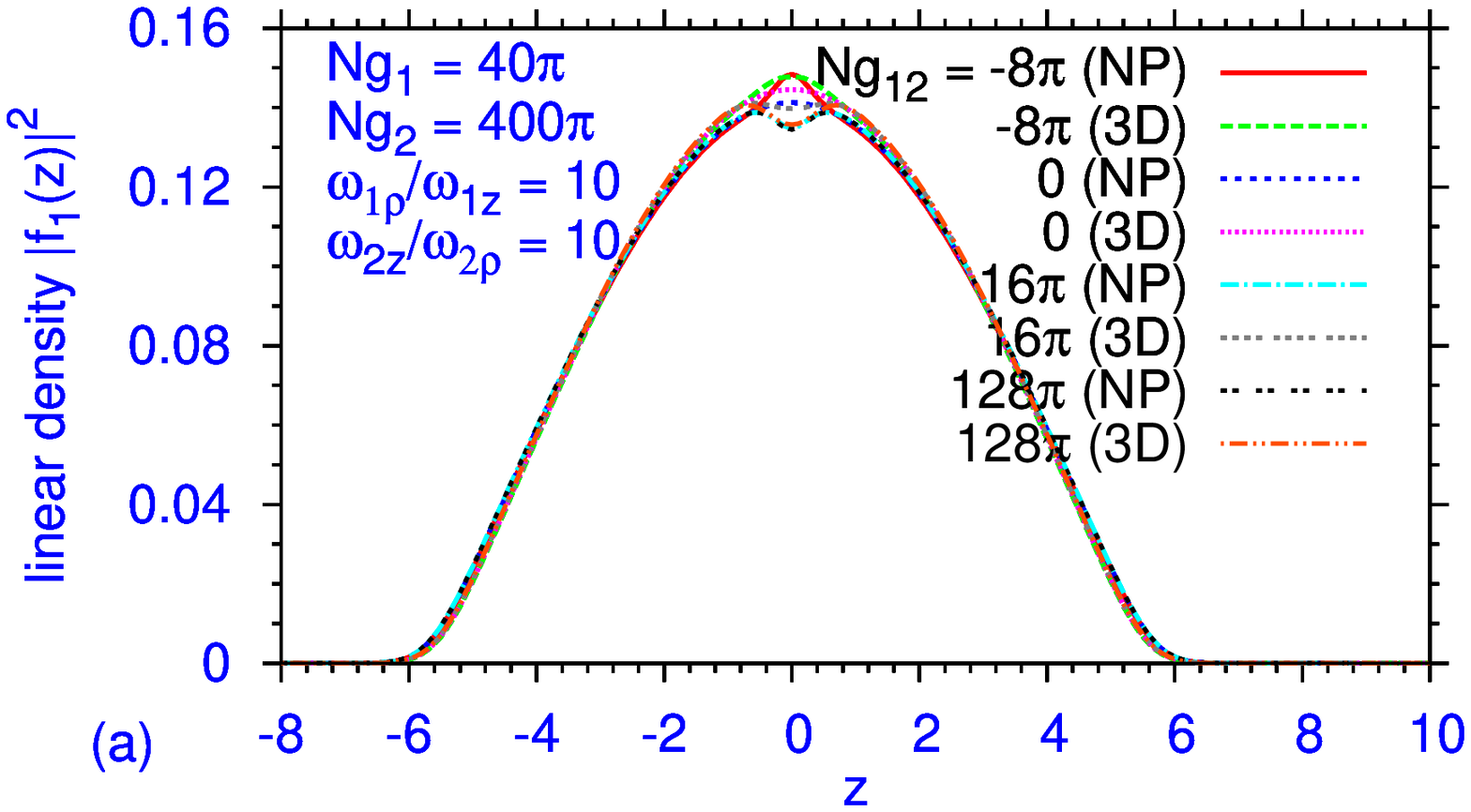}
\includegraphics[width=.7\linewidth]{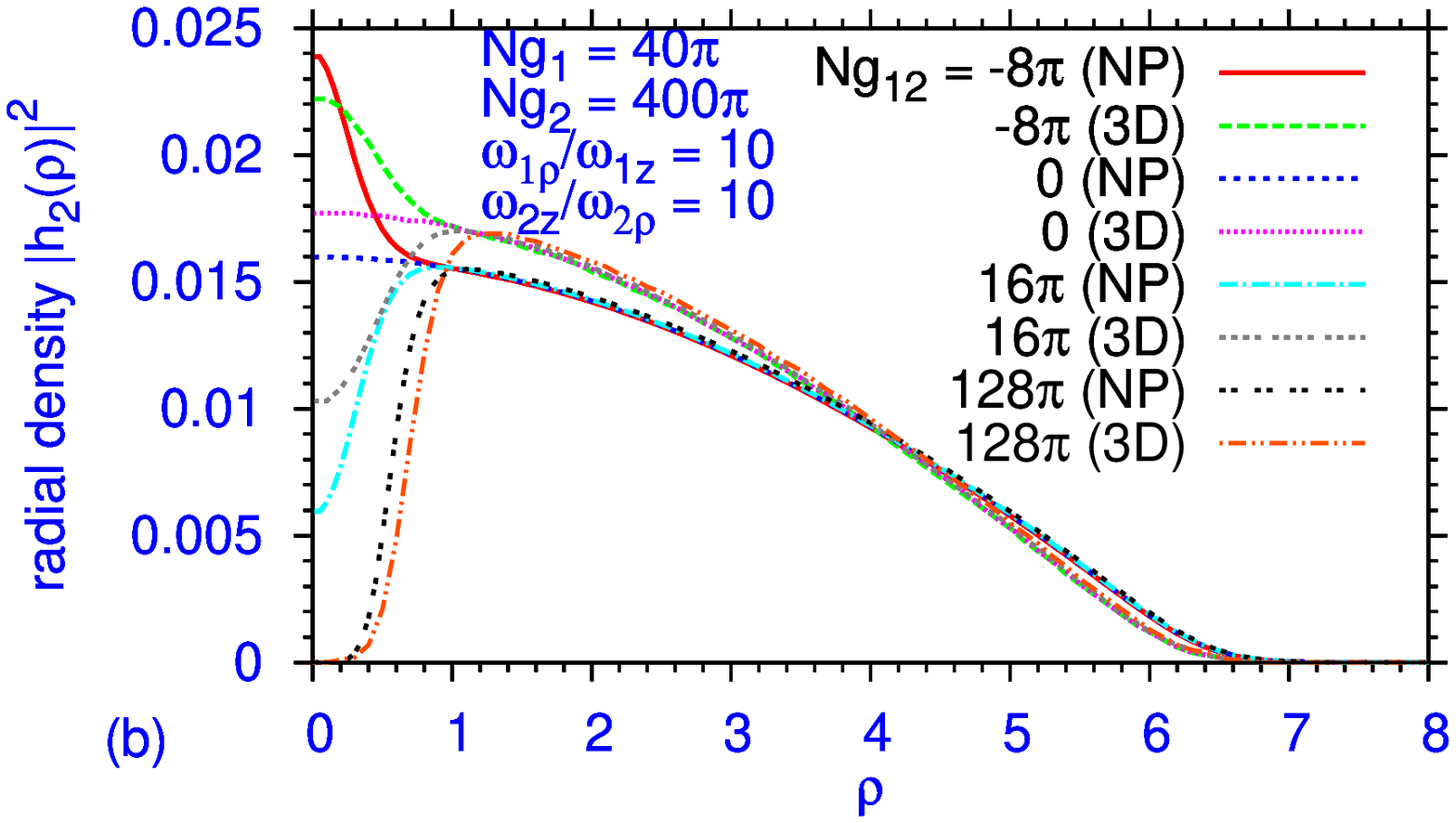}
\end{center}

\caption{(Color online) The same as in Fig. \ref{fig3} obtained using the 
reduced 
BNPSEs ({\ref{eq48}}) $-$ (\ref{eq51}) (NP).}

\label{fig4}
\end{figure}

\subsection{Stationary States}

In Figs. {\ref{fig1}} (a) and (b) we plot the linear and radial 
densities $|f_1(z)|^2$ and $h_2(\rho)|^2$, respectively,  
 obtained from the 3D equations (\ref{3}) and (\ref{4}) as well as the 
BNLSE (\ref{25}) and (\ref{26}) for $Ng_1=0$ and 
$Ng_2=40\pi$ and for different $Ng_{12}$. In this case in the absence of 
coupling ($g_{12}=0$) the two linear densities are identical and the two
radial densities 
very close to each other.  The BNLSE (\ref{25}) and (\ref{26}) 
provide a faithful representation of the densities even in the presence 
of coupling as can be seen from Fig. \ref{fig1}.

In Figs. {\ref{fig2}} (a) and (b) we plot the linear and radial 
densities $|f_1(z)|^2$ and $h_2(\rho)|^2$, respectively, as in Fig. 
{\ref{fig1}} (a) and (b), for $Ng_1=4\pi$ and $Ng_2=40\pi$. In this case 
even for $g_{12}=0$, there is some difference between the results of the 
binary 3D GPE  and the BNLSE. This difference is consistent with 
the previous studies \cite{sala1,camilo}. The agreement between the 
densities obtained from the binary 3D GPE  and the BNLSE continues to be 
good. The small discrepancy between the results of the binary 3D 
GPE  and the BNLSE 
 continues as a nonzero $g_{12}$ is introduced. This discrepancy 
does not increase as  $g_{12}$  is increased to moderate values as we see 
from Fig. \ref{fig2}.
In Figs. {\ref{fig3}} (a) and (b) we plot the linear and radial densities 
for $Ng_1=40\pi$ and
$Ng_2=400\pi$. In this case there is a larger difference between 
the results of 3D and reduced 1D-2D models. What is interesting is 
that the agreement between  the results of the two models does
not deteriorate with the increase of $g_{12}$ up to $128\pi$.

\begin{figure}
\begin{center}
\includegraphics[width=.7\linewidth]{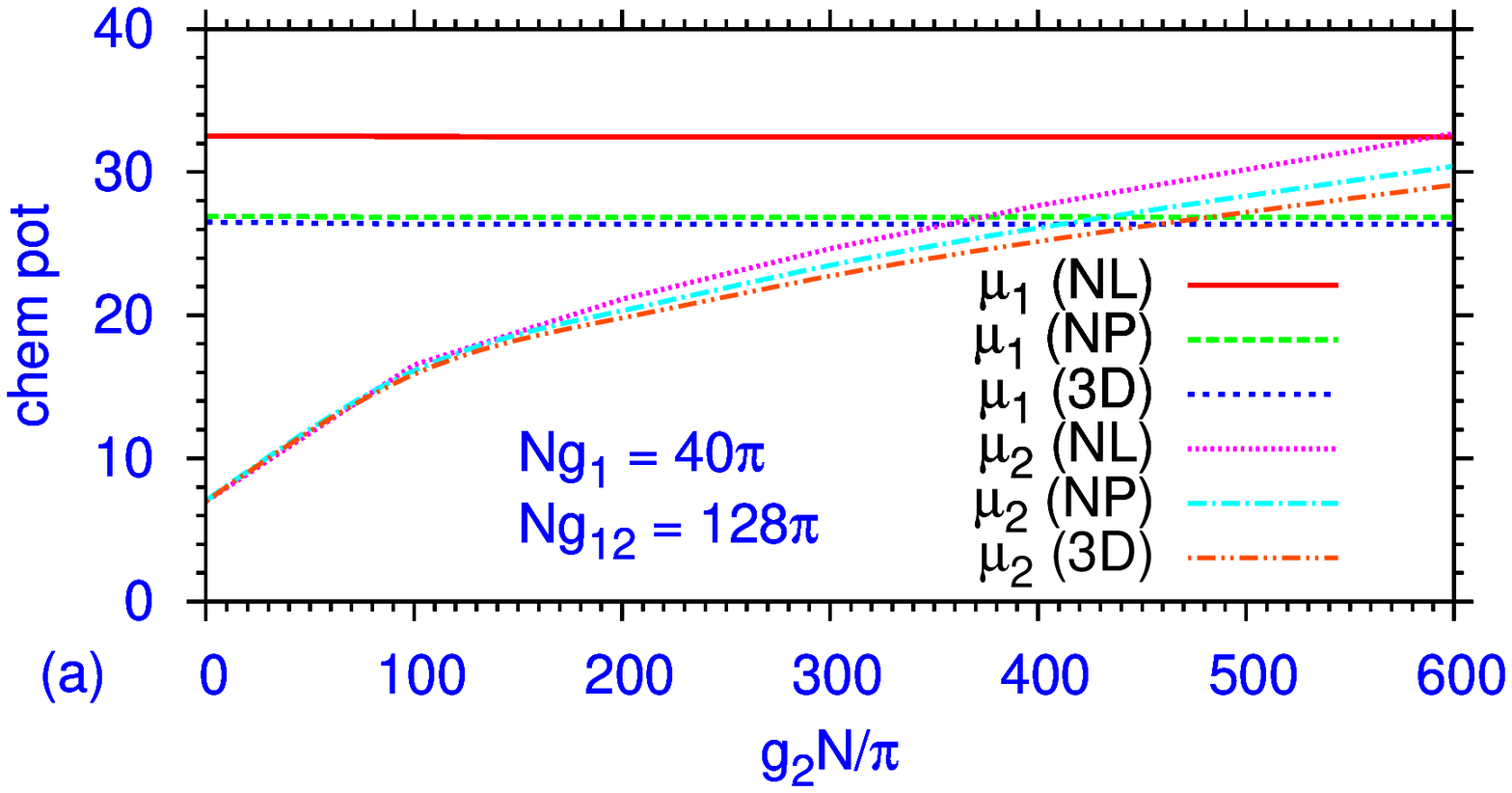}
\includegraphics[width=.7\linewidth]{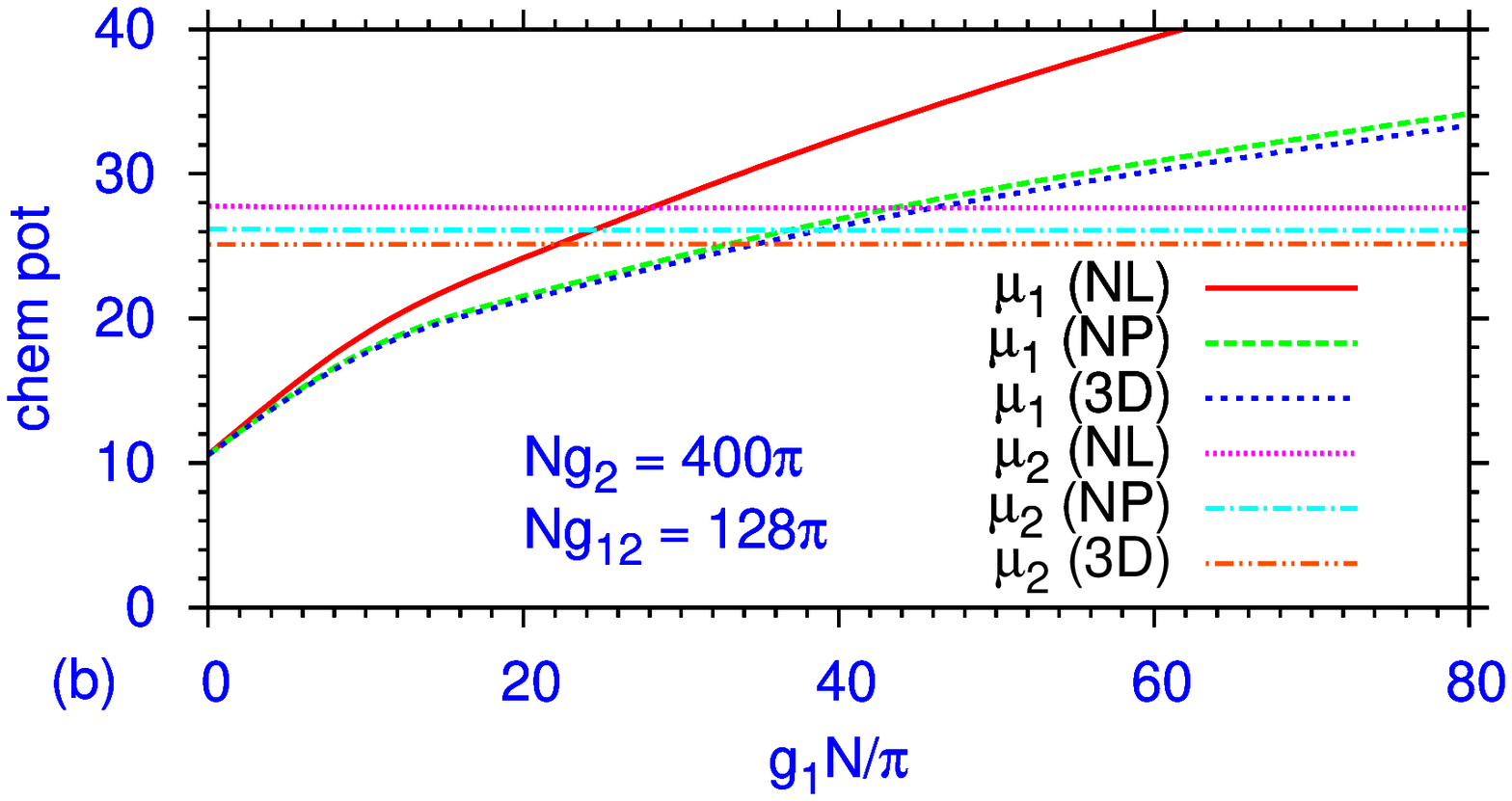}
\end{center}

\caption{(Color online)
(a) Chemical potential of the two components $\mu_1$
and $\mu_2$
calculated using the 3D
GPE (\ref{3}) and (\ref{4}) (3D), 
 the reduced BNPSEs (\ref{eq48})
 $-$ (\ref{eq51}) (NP), and 
the reduced BNLSE (\ref{25}) and (\ref{26})
(NL)
versus $g_2N/\pi$ for $Ng_1=40\pi, N_g{12}=128\pi$. 
 (b) Chemical potential of the two components $\mu_1$
and $\mu_2$
calculated using the 3D GPE
 (\ref{3}) and (\ref{4}) (3D), 
 the reduced BNPSEs (\ref{eq48}) 
$-$ (\ref{eq51}) (NP), and 
the reduced BNLSE (\ref{25}) and (\ref{26})
(NL)
versus $g_1N/\pi$ for $Ng_2=400\pi, Ng_{12}=128\pi$.
}

\label{fig5}
\end{figure}

From the results displayed in Figs. \ref{fig1} $-$ \ref{fig3} we find 
that the BNLSE (\ref{25}) and (\ref{26}) provide an 
excellent account of the actual state of affairs, when compared with the 
full 3D binary GPE (\ref{3}) and (\ref{4}), specially for small 
nonlinearities $Ng_1$ and $Ng_2$.

We next attempt an 
improvement over the  BNLSE (\ref{25}) and (\ref{26}) of Sec.
\ref{II} by applying BNPSE of Sec. \ref{III}. In other words, 
we attempt a solution 
of Eqs. (\ref{eq48}) $-$ (\ref{eq51}), in place of Eqs. (\ref{25})
and (\ref{26}), to reduce the discrepancy noted in Figs. \ref{fig3}
(a) and (b)
between the results of the 3D and BNLSE.  
The results so
obtained 
 are plotted in Figs. \ref{fig4} (a) and (b) for $Ng_1=40 \pi$ and 
$Ng_2=400\pi$, which show
a significant improvement over the results in Figs. \ref{fig3}
(a) and (b). The large discrepancy between the results of the 
3D binary GPE and the BNLSE
for the linear density of component 1  has virtually disappeared in 
Fig. \ref{fig4} (a)
and that for the radial density of component 2 is significantly 
reduced in Fig. \ref{fig4} (b).
This shows that although the results from the effective 
BNLSE of Sec.
\ref{II}  are quite good for small nonlinearities, significant 
improvement can be obtained from the BNPSE of Sec. \ref{III}, 
especially for larger values of nonlinearities.

Finally, we consider the chemical potential $\mu_1$ and $\mu_2$ of the 
two components calculated using the 3D binary GPE (\ref{3}) and (\ref{4}), 
 the BNPSE (\ref{eq48}) $-$ (\ref{eq51}) as well as the 
BNLSE (\ref{25}) $-$ (\ref{26}). The equations 
for the chemical potentials are obtained by replacing the time 
derivative $i\hbar\partial_t$ by $\mu_1$ and $\mu_2$, respectively, in 
the dynamical equations for component 1 and 2, respectively. The  
chemical potentials are plotted in Fig. \ref{fig5} (a) versus $g_1N/\pi$ 
for $Ng_2=400\pi$ and $Ng_{12}=128\pi$. In Fig. \ref{fig5} (b) we plot 
the chemical potentials versus $g_2N/\pi$ for $Ng_1=40\pi$ and 
$Ng_{12}=128\pi$.  The chemical potentials obtained from the 
BNLSE (\ref{25}) $-$ (\ref{26}) are good but those calculated from 
the BNPSE (\ref{eq48}) $-$ (\ref{eq51}) are better approximations to the 
results obtained from  the 3D binary GPE 
(\ref{3}) and (\ref{4})
in all cases.

\label{IV}

\subsection{Dynamics}

\begin{figure}
\begin{center}
\includegraphics[width=.8\linewidth]{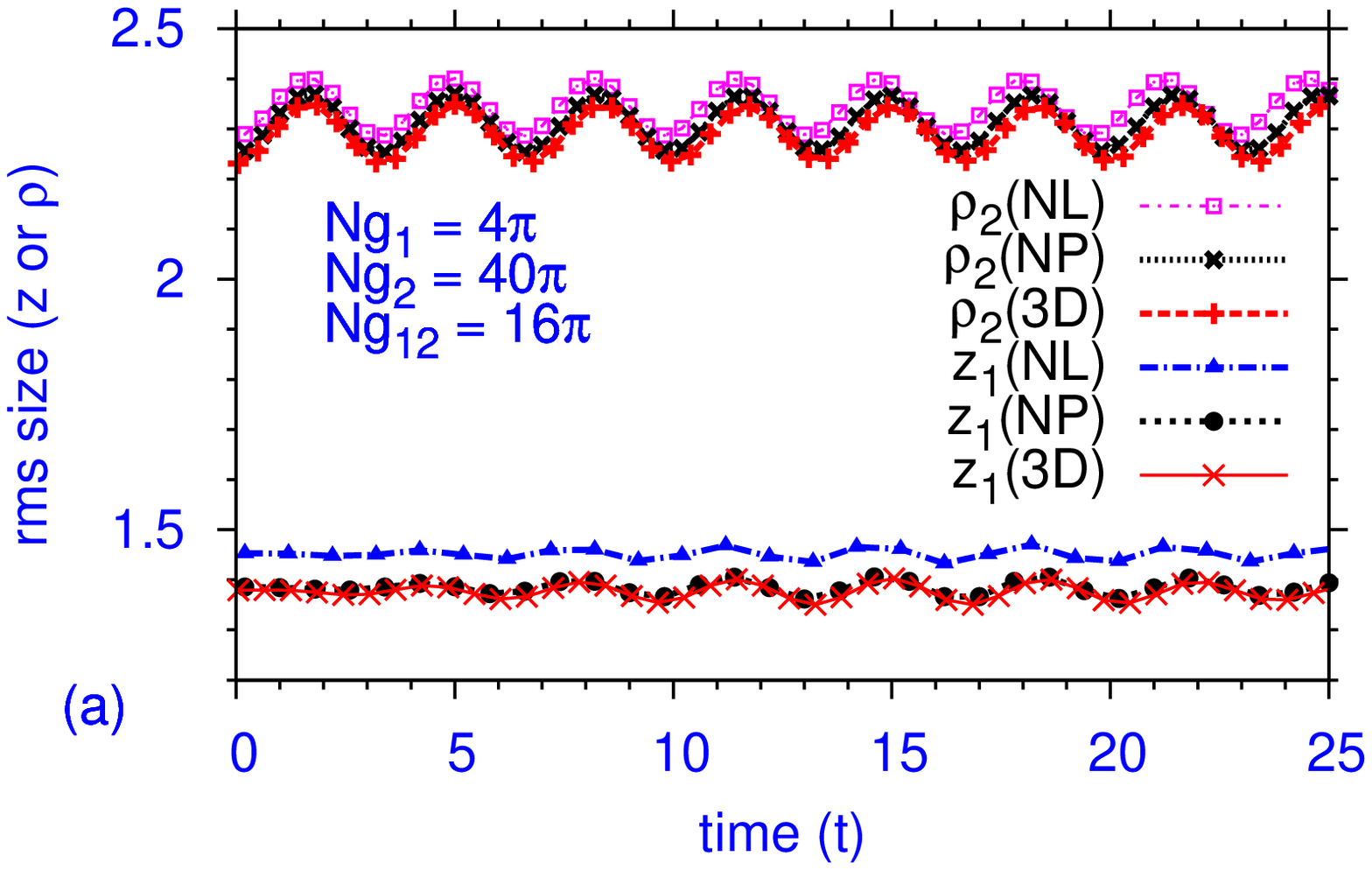}
\includegraphics[width=.8\linewidth]{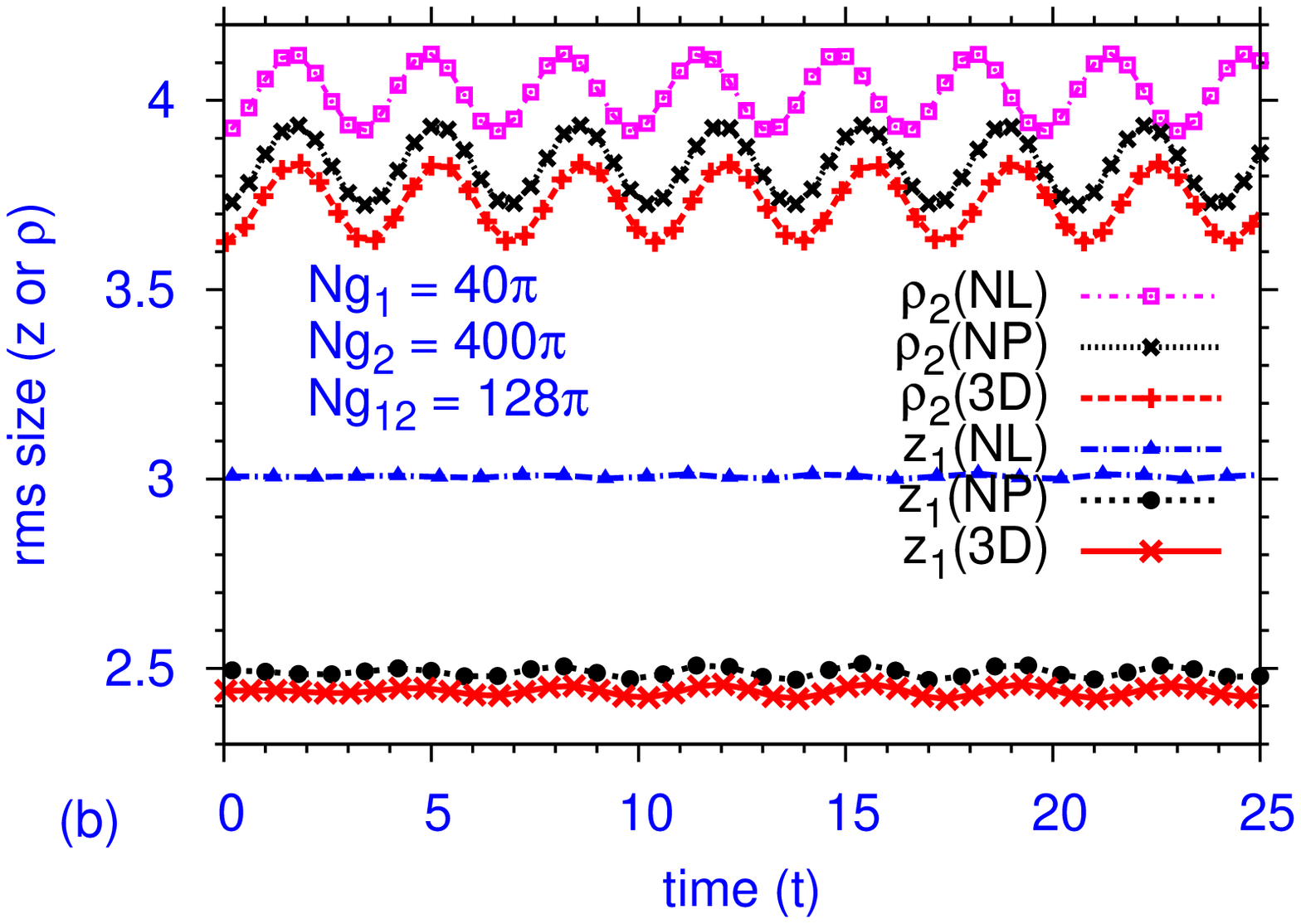}
\end{center}

\caption{(Color online) Evolution of axial and radial rms sizes of the 
cigar- and disk-shaped components of the binary BEC after a sudden 
reduction of the radial angular
frequency of the disc-shaped component by $5\%$, 
for the nonlinearities (a) $Ng_1=4\pi, Ng_2=40\pi, Ng_{12}=16\pi$ and 
(b) $Ng_1=40\pi, Ng_2=400\pi, Ng_{12}=128\pi$. The cigar-shaped BEC is 
perpendicular to the disk-shaped BEC. 
}

\label{fig6}
\end{figure}

Now we investigate if the reduced equation presented here can satisfactorily 
describe the  dynamics of the perpendicular  1D-2D system. For this purpose
first we consider the coupled system presented in Figs. \ref{fig2} (a) 
and (b) with $Ng_1=4\pi, Ng_2=40\pi,$ and $Ng_{12}=16\pi$. 
After the stationary 
condensate is created during time evolution of the numerical routine 
using real-time propagation, we reduce  the radial
angular  frequency of the disk-shaped
quasi-2D BEC suddenly by 5$\%$ (the new angular 
frequencies being $\omega_{1\rho}=
\omega_{2z}=10, \omega_{1z}=1,
\omega_{2\rho}=0.95,$)
and study the subsequent dynamical evolution of the system. The relevant 
dynamics is best illustrated by plotting the evolution of 
the rms axial size of the 
cigar-shaped quasi-1D BEC 
and 
the rms radial size of the 
disk-shaped quasi-2D BEC 
in Fig. \ref{fig6} (a) and (b) for two sets of nonlinearities
exhibited in Figs. \ref{fig2} and \ref{fig3}. 
Here we show the results of the 3D  GPE
(\ref{3}) and (\ref{4}) (3D),
 the reduced BNPSEs (\ref{eq48})
 $-$ (\ref{eq51}) (NP), and
the reduced BNLSE (\ref{25}) and (\ref{26})
(NL). For both sets of nonlinearities illustrated in Figs. \ref{fig6}
(a) $Ng_1=4\pi, Ng_2=40\pi,Ng_{12}
=16\pi$ and (b) $Ng_1=40\pi, Ng_2=400\pi,Ng_{12}
=128\pi$, the NP reduction produced  far better results compared to 
the NL reduction. In both cases the angular
frequency of the quasi-2D disk-shaped 
BEC was modified and both the NL and NP schemes as well as the full 3D 
calculation exhibited sinusoidal oscillation for the axial size of the 
disk-shaped BEC right from $t=0$. It took some five units of time for 
similar sinusoidal  oscillation to initiate in the unperturbed cigar-shaped 
component due to the non-zero coupling term  $g_{12}$. The oscillations 
in rms sizes in the NP scheme were always in phase with 
those of the full 3D calculation. But for larger nonlinearity and at 
large times the oscillations 
in rms sizes in the NL scheme are not always in phase with those 
of the full 3D calculation. 
The differences between the  rms sizes of
the NL and NP schemes compared to the full 3D calculation were consequences 
of the same differences, existing at $t=0$,  in the corresponding 
stationary results in the 
absence of any perturbation in the 
angular frequency.

\section{SUMMARY}
\label{V}

In this paper we derived and studied simple binary  reduced non-polynomial 
Schr\"odinger equations (BNPSEs) for the description of a binary BEC where 
the two components subject to distinct trap symmetries
belong to two different spatial dimensions. We considered three possibilities: 
 the first component is in 3D and the second in (a) 1D (cigar shape) or 
(b) 2D (disk shape), and (c) the first component is in 1D and the 
second in 2D. The 1D and 2D configurations were achieved in an 
axially-symmetric setting with a strong harmonic trap in the transverse 
radial and axial directions, respectively. The BNPSEs were obtained by a 
Lagrangian variational scheme after approximating the spatial 
wave-function component in the transverse direction by a generic 
Gaussian function. A simpler set of reduced equations, the 
BNLSE  with power-law 
nonlinearity,  were obtained with a simple ansatz for the 3D wave function 
where the wave-function component in the transverse direction is taken 
to be the ground state in the corresponding harmonic trap. 
 
We tested the accuracy of the reduction scheme above by numerically 
solving the 3D binary GPE in the coupled 1D-2D case where the cigar-shaped 
component is in the $z$ direction and the disk-shaped component lie 
in the $(x,y)$ plane. We calculated the density profiles and the chemical 
potentials of the two components. For smaller values of inter- and 
intra-species nonlinearities, the BNLSE (\ref{25}) and 
(\ref{26}) with power-law nonlinearity produced results for density and 
chemical potential in good agreement with the binary 3D GPE. For 
larger nonlinearities the BNPSE (\ref{eq48}) $-$ (\ref{eq51}) produced 
significant improvement over the BNLSE (\ref{25}) and 
(\ref{26}). We also studied small oscillation of the binary 1D-2D 
BEC, initiated by a sudden change of the radial 
angular frequency of the 2D
component. For larger nonlinearities and large times, the BNPSE (\ref{eq48}) 
$-$ (\ref{eq51}) produced better results for dynamics 
compared with the  BNLSE (\ref{25}) and
(\ref{26}).
To conclude, the  BNLSE  (\ref{25}) and (\ref{26}) as 
well as the BNPSE (\ref{eq48}) $-$ (\ref{eq51}) are shown to be very 
useful to investigate a binary 1D-2D BEC, where the cigar-shaped BEC is 
perpendicular to the disk-shaped one. Similar reduced equations are 
derived for other symmetries, such as the 1D-3D and 2D-3D cases as well 
as the 1D-2D case where the cigar-shaped BEC is in the plane of the 
disk-shaped one.

\acknowledgments

CNPq and FAPESP (Brazil) provided partial support.

\end{document}